\documentclass[fleqn,usenatbib]{mnras}
\usepackage{fix-cm}

\usepackage{newtxtext,newtxmath}
\RequirePackage[fixed]{fontawesome5}

\usepackage[T1]{fontenc}

\DeclareRobustCommand{\VAN}[3]{#2}
\let\VANthebibliography\thebibliography
\def\thebibliography{\DeclareRobustCommand{\VAN}[3]{##3}\VANthebibliography}

\usepackage{graphicx}	
\usepackage{amsmath}	
\usepackage{algpseudocode}
\usepackage{algorithm}
\usepackage{float}
\usepackage{tikz,xcolor,hyperref}
\definecolor{lime}{HTML}{A6CE39}
\DeclareRobustCommand{\orcidicon}{
\begin{tikzpicture}
\draw[lime, fill=lime] (0,0)
circle[radius=0.16]
node[white]{{\fontfamily{qag}\selectfont \tiny \.{I}D}}; 
\end{tikzpicture}
\hspace{-2mm}
}
\usepackage{listings}

\title[Capivara: A Spectral-based Segmentation Method for IFU Data Cubes]{
\includegraphics[scale=0.025]{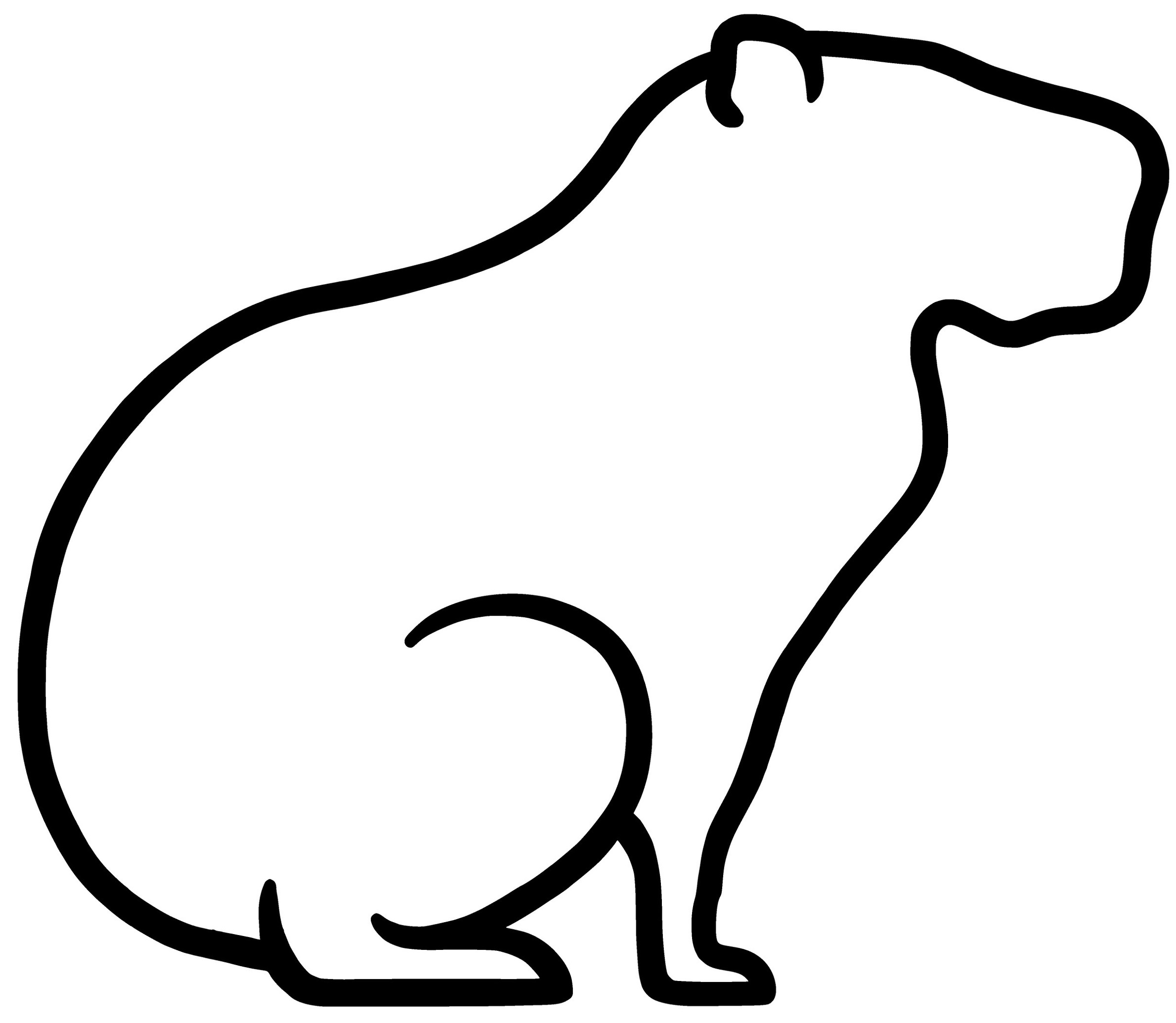}
Capivara: A Spectral-based Segmentation Method for IFU Data Cubes}

\DeclareRobustCommand{\orcidicon}{
\begin{tikzpicture}
\draw[lime, fill=lime] (0,0)
circle[radius=0.16]
node[white]{{\fontfamily{qag}\selectfont \tiny \.{I}D}}; 
\end{tikzpicture}
\hspace{-2mm}
}
\foreach \x in {A, ..., Z}{%
\expandafter\xdef\csname orcid\x\endcsname{\noexpand\href{https://orcid.org/\csname orcidauthor\x\endcsname}{\noexpand\orcidicon}}
}

\author[Rafael S. de Souza]{Rafael S. de Souza\hspace{-1.5mm}\orcidA{},$^{1,2,3}$\thanks{E-mail: \href{mailto:r.da-silva-de-souza@herts.ac.uk}{r.da-silva-de-souza@herts.ac.uk}}
Luis G. Dahmer-Hahn\hspace{-1.5mm}\orcidH{},$^{4}$\thanks{E-mail: \href{mailto:luisgdh@gmail.com}{luisgdh@gmail.com}
}
Shiyin Shen\hspace{-1.5mm}\orcidB{},$^{4}$
Ana L. Chies-Santos\hspace{-1.5mm}\orcidC{},$^{5}$
\newauthor
Mi Chen\hspace{-1.5mm}\orcidG{},$^{4}$
P. T. Rahna\hspace{-1.5mm}\orcidR{} $^{6}$
Paula Coelho\hspace{-1.5mm}\orcidP{} $^{2}$
Rogério Riffel\hspace{-1.5mm}\orcidK{} $^{5}$
Renhao Ye\hspace{-1.5mm}\orcidE{},$^{4}$ 
Behzad Tahmasebzadeh\hspace{-1.5mm}\orcidF{} $^{7}$
\\
$^{1}$ Centre for Astrophysics Research, University of Hertfordshire, College Lane, Hatfield, AL10~9AB, UK \\
$^{2}$ Instituto de Astronomia, Geofísica e Ciências Atmosféricas, USP, Rua do Matão 1226, 05508-090, São Paulo, Brazil\\
$^{3}$ Department of Physics \& Astronomy, University of North Carolina at Chapel Hill, NC 27599-3255, USA\\
$^{4}$ Shanghai Astronomical Observatory, Chinese Academy of Sciences, 80 Nandan Rd., Shanghai 200030, China\\
$^{5}$Instituto de Física 
Universidade Federal do Rio Grande do Sul 
Av. Bento Gonçalves 9500 
Porto Alegre, R.S. 90040-060, Brazil\\
$^{6}$ Centro de Estudios de Física del Cosmos de Aragón (CEFCA), Plaza San Juan 1, 44001 Teruel, Spain\\
$^{7}$ Department of Astronomy, University of Michigan, 
1085 S. University Ave., 
Ann Arbor, MI 48109, USA
}

\date{Accepted XXX. Received YYY; in original form ZZZ}

\pubyear{2015}

\begin{document}
\label{firstpage}
\pagerange{\pageref{firstpage}--\pageref{lastpage}}
\maketitle

\begin{abstract}
We present {\sc capivara}, a fast and scalable spectral-based segmentation package designed to study astrophysical properties within distinct structural components of galaxies. This  spectro-segmentation code for integral field unit (IFU) data provides a holistic view of galactic structure, moving beyond conventional radial gradients and the bulge-plus-disk dichotomy. 
It enables detailed comparisons of stellar ages and metallicities across components, and naturally identifies outliers by grouping spaxels according to dominant spectral features. 
The algorithm leverages Torch's scalability and GPU acceleration, outputting a masked FITS file that assigns each pixel to its respective group and generates the corresponding one-dimensional spectrum per group, without relying on Voronoi binning. We demonstrate the capabilities of the method using a sample of MaNGA galaxies, combining \textsc{capivara} segmentation with the \textsc{starlight} spectral fitting code to derive stellar population and ionized gas properties. 
The method effectively identifies regions with similar spectral properties across both continuum and emission lines.  By aggregating the spectra of these regions, we enhance the signal-to-noise ratio of the analysis while preserving the spectral coherence within each group.{\sc capivara} is released under an  MIT license and is available at \href{https://github.com/RafaelSdeSouza/capivara}{\faGithub}.
\end{abstract}




\section{Introduction}

Galaxies, the fundamental building blocks of the cosmos, comprise a rich diversity of stellar components, including bulges, disks, bars, spiral arms, and nuclear star clusters. Each of these structures exhibits distinct signatures in age, metallicity, and kinematics, shaped by the interplay of galactic formation and evolution over cosmic time \citep[e.g.,][]{Hubble1926,Cappellari2016,tabor17,coccato18,sanchez2020,johnston22,Haussler+22,nedkova24,Jegatheesan24,limadias24,Zanatta2024}. The bulge of the Milky Way primarily consists of an ancient population over 10 billion years old and rich in metals, but it also contains intermediate-age, thick-disk-like stars and a relatively young nuclear disk. This contrasts with its thin disk, which exhibits metallicity closer to that of the Sun and a considerably younger average age \citep{hayden2015chemical,Howes2015, ness2016metallicity, chiti2021metal}.

Upon detailed analysis, the Milky Way reveals thick and thin disks, a bar, spiral arms, a nuclear disk, and a complex halo \citep[e.g.,][]{anders2014chemodynamics, helmi2020streams, Queiroz2021, ardern2024pristine}. Other models have been suggested to explain the formation of the Galactic disk, involving processes such as star formation, gas accretion, galaxy mergers, and stellar migration \citep{di2016disc, 2023ApJ...954..124I}.  Unlike our Galaxy, such detailed scrutiny of structural components in galaxies beyond the Local Group is not possible, often leading to crude divisions between disk and bulge components \citep{buitrago08, nedkova24}, although advances have been achieved for nearby galaxies with newer instrumentation \citep{timer19, GECKOS24}. Yet, the distinction between bulge and disk populations is pivotal for understanding the processes behind galaxy formation and evolution. It is generally accepted that massive galaxies form hierarchically in a simplified two-phase process: a rapid dissipative phase (which includes a starburst phase), followed by multiple mergers \citep{Whitney2021}. The James Webb Space Telescope is showing that massive galaxies are already in place at $z \sim 3$ \citep{carnall23}. Bulges are typically older and more metal-rich relative to disks \citep[][]{Lah2023}. The influence of bulges and disks on galaxy evolution depends on factors like galaxy/halo mass, environment, and redshift \citep{Conselice14,Cimatti+19}. 

A galaxy's environment \citep{peng2010b} also influences its properties \citep{lani13,Lima2021}, with central massive galaxies typically growing inside-out due to mechanisms like gas accretion and Active Galactic Nucleus (AGN) feedback, which predominantly impact central regions \citep[e.g.,][]{Croton2006,Goddard2017,kim2011galaxy, Krumholz2018, baker2023inside}. Conversely, satellite galaxies often exhibit outside-in profiles due to external quenching forces that suppress star formation from the edges inward, a phenomenon underscored by mechanisms such as major mergers \citep{lin2019sdss, Cortese2021, samuel2022extinguishing}. Recent findings of outside-in radial gradients in distant, low-mass satellites \citep{Sandles2023} emphasize environmental quenching's role from as early as $z = 2$. Alongside stellar histories, gas distribution and properties offer insights into quenching processes, with internal mechanisms leading to central outflows and environmental quenching leaving extended shocked gas distributions, identifiable by specific emission line ratios \citep{Carniani2023}.


The use of Integral Field Units (IFUs) has enabled more detailed studies of galaxy properties by providing simultaneous spectral and spatial information, allowing for the examination of chemical abundances, kinematics, and star formation histories in a spatially resolved manner. These observations are an integral part of numerous low- and high-redshift surveys, including SAURON \citep{Bacon2001}, the Spectroscopic Imaging Survey in the Near-infrared with SINFONI \citep[SINS;][]{Schreiber2009}, Mapping Nearby Galaxies at Apache Point Observatory \citep[MaNGA;][]{Bundy2015}, Calar Alto Legacy Integral Field Area \citep[CALIFA;][]{Sanchez2012}, the Sydney Australian Astronomical Observatory Multi-Object Integral Field Spectrograph \citep[SAMI;][]{SAMISurvey_2012, SAMIDR3_2021} Galaxy Survey, the Gemini Near-Infrared Field Spectrograph \citep[NIFS;][]{Riffel2018MNRAS}, Time Inference with MUSE in Extragalactic Rings \citep[TIMER;][]{timer19}, and the Middle Ages Galaxy Properties with Integral Field Spectroscopy \citep[MAGPI;][]{Foster2021PASA}. 

IFU data have been used in morphological decomposition to identify spaxels belonging to bulge- or disk-dominated regions (and other structures), enabling the examination of the properties of the stellar populations in these regions. For example, \citet{Wisotzki2003} used 2D modeling techniques on IFU data to deblend the quadruple QSO and gravitational lens HE 0435-1223. A similar approach was used to separate the nuclear and host galaxy of the central region of 3C 120 \citep{Sanchez2004} and to extract the deblended spectra of the galaxies in the core of Abell 2218 \citep{Sanchez2007}. A noteworthy package is the Bulge-Disk Decomposition of IFU data \citep[BUDDI;][]{Johnston2017}, which utilizes {\sc GalfitM} \citep{Haussler2013}, a modified version of {\sc Galfit} \citep{Peng2002, Peng2010a}, to model the light profile of multi-wavelength images of a galaxy. \citet{Mendez2019} employed C2D, a multi-component decomposition, which consists of fitting a galaxy's surface-brightness distribution at each wavelength (quasi-monochromatic image) via a 2D photometric decomposition code. \citet{bittner+21} studies the kinematic and stellar population of inner bars in three disk galaxies, based on the multi-component photometric decompositions from \citet{timer_doublebars19}.

One shortcoming of current image decomposition methods is their reliance on preset categories—such as halos, bulges, disks, bars, and rings—which exclude structures that do not fit, like asymmetric features. Additionally, when working with IFUs, these methods typically treat each wavelength independently rather than as part of a high-dimensional tensor, potentially overlooking important correlations—such as those between bluer wavelengths and \ion{H}{II} emission. This results in the loss of useful information that could provide insights into the physical mechanisms behind these features. 

This work explores an alternative approach for spectral categorization of different physical regions in galaxies. Instead of using preconceived two-dimensional profile models, our method employs a    spectral-based segmentation approach solely relying on {\it spectral similarity}. This allows for greater flexibility in characterizing complex structures, such as those found in mergers and irregular galaxies. By utilizing {\it unsupervised hierarchical clustering}, our approach directly targets the wavelength space instead of treating the data cube as a stack of images at each wavelength. This methodology leverages \texttt{torch} \citep{paszke2019pytorch}, a deep learning library designed for tensor computations and GPU acceleration, enabling fast and efficient matrix calculations. Its capabilities make it particularly suitable for the  segmentation of large datasets, offering a powerful tool for the analysis of IFU data cubes.

This paper is organized as follows: In Section \ref{s:decomposition}, we provide a detailed description of our method. Section \ref{sec:data} briefly describes the data used in our study. In Section \ref{sec:app}, we demonstrate the practical application of our code on five MaNGA galaxies. We compare our approach with existing methods from the literature and present our conclusions in Section \ref{sec:voronoi}.

\section{Method: Spectro-Segmentation}
\label{s:decomposition}

Our model classifies each IFU pixel \(\phi = \left\{ \phi_i \in \mathbb{R}^{x \times y \times w} \right\}_{i=1}^{N}\) into \(C\) clusters, ensuring that spectra within a cluster are similar, while those in different clusters are dissimilar. Here, \(N\) denotes the total number of pixels $\left(x \times y\right)$, and \(w\) is the number of spectral wavelengths.  We measure similarity using the \(\ell_2\) distance metric, given by:
\begin{equation}
\ell_p(u, v) = \left( \sum_{i=1}^d | u_i - v_i |^p \right)^{\frac{1}{p}}, 
\label{eq:l1}
\end{equation}
where $p=2$ represents the Euclidean distance. Here,  $u$ and $v$ represent two independent vectors of features, $i$ is the index of the feature,  and $d$ is the number of features. 

In our case, the features are the flux values at each wavelength bin of a spaxel in the IFU. 
While other possible choices exist, such as the Kullback–Leibler divergence \citep{Kullback1951}, mutual information \citep{Li1990}, and so forth, we opt for the $\ell_2$ norm due to its simplicity and its relation to $\chi^2$ statistics, a standard method for spectral fitting and template matching. This approach preserves the common underlying assumption that spectra can be compared via their respective Euclidean distances. The properties of the spectra include their overall shape, the strength of particular spectral lines, and other distinguishing characteristics that can influence the similarity measure.

\begin{figure*}
    \includegraphics[width=\textwidth]{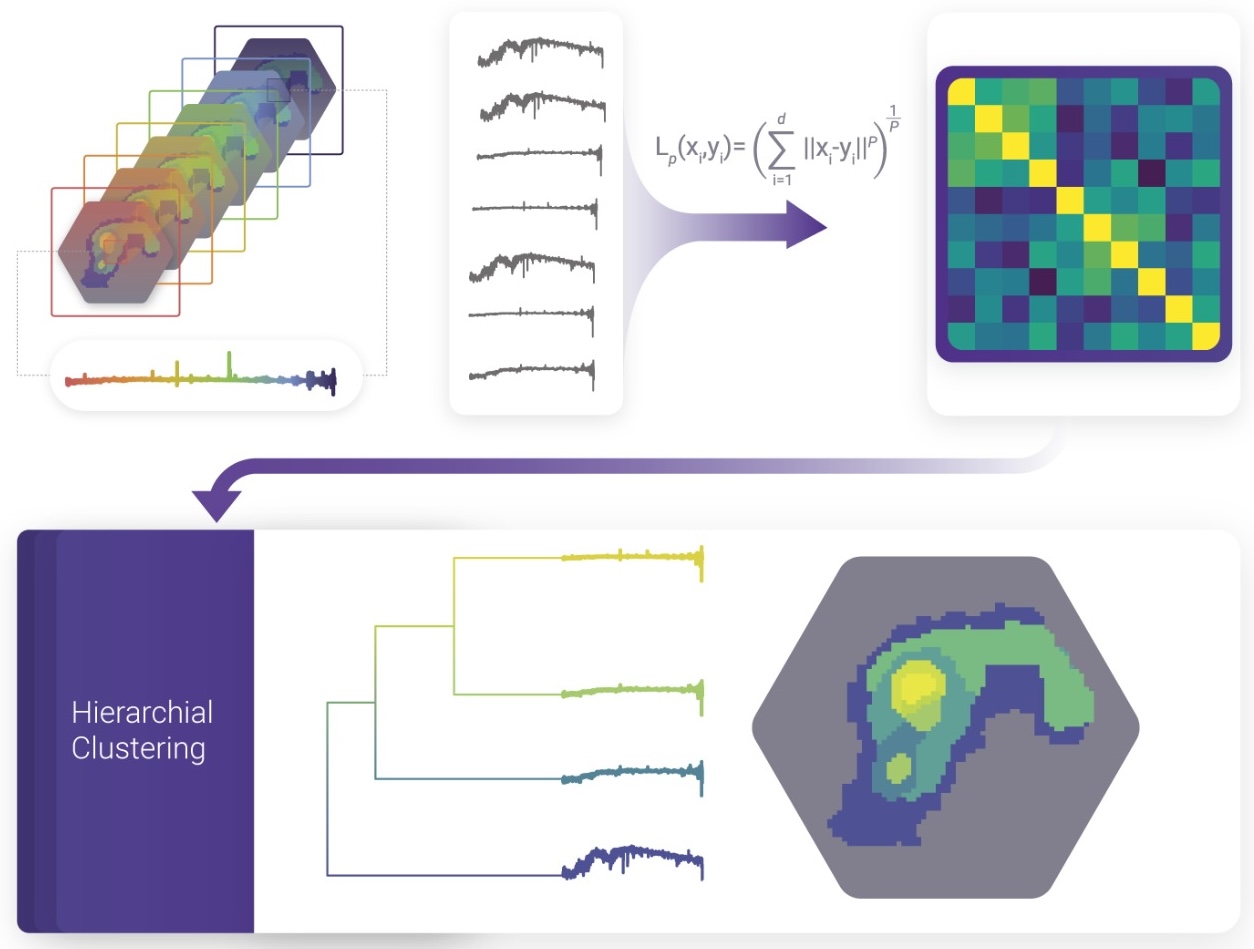}
    \caption{Workflow illustration of our spectro-based segmentation method. The model reads an IFU data cube composed of multiple wavelength channels, each represented by a slice in the left image. Hierarchical clustering is then performed on the dissimilarity matrix computed from pairwise distances between all spectra within the cube. Once the groups are assigned, the data is back-transformed into a 2D matrix where each group represents spectra with similar features. This process identifies galactic structures through spectral characteristics, enabling a refined structural decomposition.}
    \label{fig:schematics}
\end{figure*}

Once the similarity measure, in the form of a distance matrix, is established, the algorithm begins by assigning each spectrum from a given spaxel to its own group. This initial step ensures that each spectrum is considered a unique entity, facilitating a more granular analysis. The algorithm then proceeds iteratively, merging the two most similar clusters in each iteration until a predetermined number of groups is achieved. This process of successive merging can be visualized as the construction of a hierarchical tree, where groups formed by merging similar spectra are placed lower on the tree, and those that are less similar are positioned higher.

One of the significant advantages of hierarchical clustering is its ability to detect associations at various levels of granularity \citep[e.g.][]{Murtagh2014,rafael2023}. For example, to identify broad categories of galaxy spectra like star-forming and passive galaxies, the algorithm can cut the tree at a higher level in the hierarchy, leading to fewer, broader groups. Conversely, for more specific distinctions within a category of galaxy spectra, we can cut the tree at a lower level, resulting in finer-grained groups. This approach allows for a hierarchical tree with broad categories at the higher levels and more specific subcategories at the lower levels. We note that during the writing of our work, independent groups have also pursued the idea of an unsupervised approach to cluster regions of similar spectra, but using a different methodology based on the Fisher Expectation-Maximization algorithm \citep{Chambon2024}.

The general workflow of our approach is depicted in Fig.~\ref{fig:schematics}. Consider a hypothetical data cube with dimensions $64 \times 64 \times 1,\!000$, where the first two dimensions represent the spatial resolution in pixels, and the last dimension represents the number of sampled wavelengths in the spectra. The $64 \times 64$ spectra are transformed into a matrix with $1,\!000$ columns and $64 \times 64$ rows, where each row represents a particular spaxel and each column represents a wavelength. On this matrix, we perform a pairwise distance estimate, which yields a dissimilarity matrix of size $4,\!096 \times 4,\!096$, where each cell represents the dissimilarity between pairs of spaxels. We then apply the hierarchical clustering model on the dissimilarity matrix, assigning each set of spectra to a particular group. Once this procedure is completed, we back-transform the list of spectral-based groups into their respective original spatial positions. The output will then be a $64 \times 64$ FITS file with an assigned group for each matrix element, which is depicted by the rightmost panel of Fig.~\ref{fig:schematics}. 

To decide the number of groups, we suggest a heuristic approach. If the goal is to separate the galaxy into bulge and disk components, two groups are often sufficient, assuming we disregard the group corresponding to the sky background. However, if the aim is to detect regions with different star-formation histories or ionization mechanisms, an increasing number of components should be used. If the objective is to increase the signal-to-noise ratio without losing information, the number of components must account for the spatial resolution, bin size, and field of view of the data. Therefore, we recommend using a sufficiently large number of groups that remain manageable for running a spectral fitting code of choice.

\section{MaNGA IFU Data}
\label{sec:data}

To showcase our approach, we use data from the MaNGA IFU survey---a major observational campaign of the Sloan Digital Sky Survey IV \citep[SDSS-IV;][]{Blanton+17} that aimed to study the internal kinematics and chemical properties of a large sample of nearby galaxies. It is the largest IFU survey of nearby galaxies to date, using the BOSS spectrograph to obtain spatially resolved spectroscopic data for approximately 10,000 galaxies in the redshift range $0.01 < z < 0.15$ (typical $z \sim 0.03$). The survey uses custom fiber bundles with a core diameter of $2\farcs0$ and IFUs ranging from 19 to 127 fibers, providing a field of view from $12\farcs0$ to $32\farcs0$. The spectra have a wavelength coverage of 3,622--10,354\,\AA\ and a spectral resolution of $R \sim 2,000$. MaNGA obtains three dithered exposures for each target galaxy. The data cube has a spatial scale of 0.5 arcseconds per pixel and varies in the number of wavelength channels based on the sampling method used. The MaNGA data cube contains 4,573 channels for logarithmic sampling and 6,732 channels for linear sampling \citep{Law_2016}.

\section{Analysis of five MaNGA galaxies}
\label{sec:app}
We demonstrate the feasibility of our method for identifying regions with different spectral properties within an extended object by selecting a sample of five morphologically diverse galaxies from MaNGA. The main properties of each galaxy, as listed in Table~\ref{tab:sample}, were obtained from NED\footnote{\url{http://ned.ipac.caltech.edu/}} and Marvin\footnote{\url{https://dr17.sdss.org/marvin/}} \citep{Aguado2019ApJS}. Our sample includes spiral, elliptical, and merger galaxies, as well as various features such as bars, rings, and individual stars in the field.

\begin{table*}
    \centering
    \begin{tabular}{lcccccc}
    MaNGA ID & Name & RA & Dec & z & Type & Diameter [kpc]\\
    \hline
    7443-12703  & VV 705  (IRAS F15163+4255) & 229.525 & 42.745 & 0.0403 & Merger & 47.91\\
    8135-12701  & UGC 3907         & 113.472 & 37.025 & 0.0618 & SABc   & 111.32\\
    8443-6102  & UGC 8730         & 207.061 & 24.777 & 0.0274 & (R)SB0 & 45.49\\
    10224-6104 & MCG +00-07-007   & 35.624 & 0.383 & 0.0248 & Spiral & 17.85\\
    11749-12701 & MCG+05-22-014   & 138.318 & 31.358 & 0.0417 & Sc     & 59.80 \\
    \end{tabular}
    \caption{Basic properties of our galaxy sample.}
    \label{tab:sample}
\end{table*}

\begin{figure*}
    \centering
    \includegraphics[width=\textwidth]{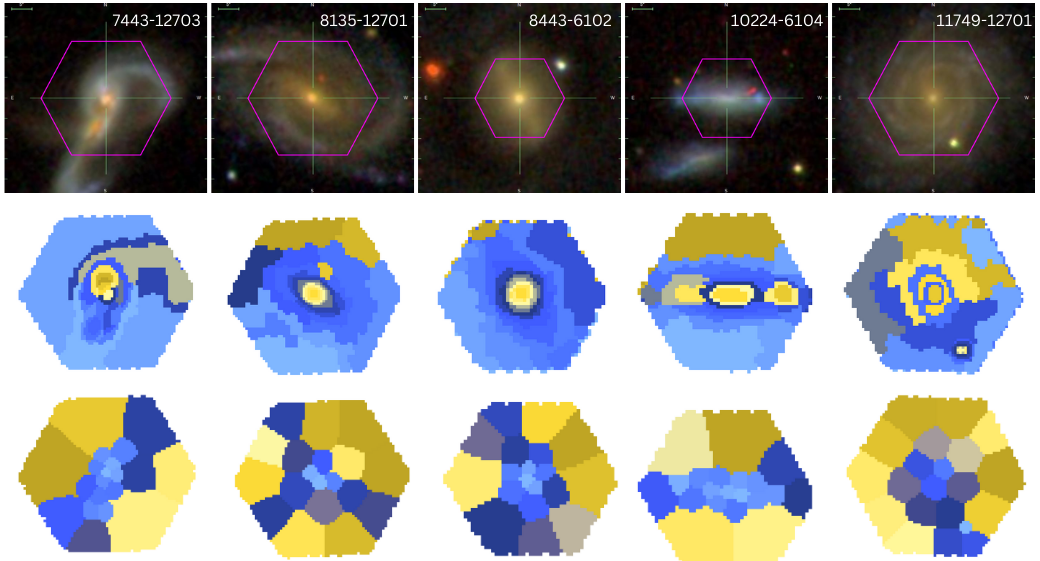}
    \caption{ \textbf{Top row}: SDSS images of five galaxies in our sample, with the MaNGA plate numbers indicated in the upper-right corner of each image. The field of view of the IFU (Integral Field Unit) is overlaid in purple. \textbf{Middle row}: Spatial distribution maps of 20 components detected by {\sc capivara} in the same galaxies. 
     \textbf{Bottom row}: Spatial distribution maps for 20 components detected by the {\sc vorbin} package.}
    \label{fig:capivara}
\end{figure*}

After applying the {\sc capivara} algorithm to the MaNGA data cubes, we obtained segmentation matrices assigning each pixel of the galaxy images to their respective component groups. Fig.~\ref{fig:capivara} presents SDSS $gri$ composite images of five selected galaxies from our sample in the top row, with their MaNGA plate numbers indicated in the upper-right corner and the corresponding IFU fields of view outlined in purple. The middle row displays the spatial distributions of the 20 components detected by {\sc capivara}, while the bottom row shows the spatial distributions obtained using the {\sc vorbin} package for the same galaxies. Comparing the segmentation maps from both approaches provides insight into their sensitivity in capturing distinct morphological and structural features.  The color schemes employed are qualitative, as the components represent discrete groups without an implied continuous ordering.\footnote{The color palette employed was inspired by Vincent van Gogh's painting \emph{The Starry Night} \citep{inla2019MNRAS}.}

A visual inspection of these figures reveals several interesting features. For example, our method effectively identifies the merger structure directly from the data cubes for the galaxy 7443-12703 (VV 705), a binary merger system in the constellation of Boötes \citep{Rupke2013, Larson2016}, with a nuclear separation of approximately 6--7,kpc and featuring long tidal tails ($\sim$40\arcsec). 

In the case of the spiral galaxies UGC 3907 (8135-12701) and MCG+05-22-014 (11749-12701), located in the constellations of Lynx and Cancer, respectively, our approach isolates what is most likely a foreground star, eliminating the need to mask the cube before further analysis, as is standard procedure in the Data Analysis Pipeline for MaNGA \citep{Westfall2019}. For the spiral galaxy MCG+00-07-007 (10224-6104), located in the constellation Cetus, the model detects at least three distinct spots, which we will further discuss as representing different stellar populations. These results illustrate the method’s efficiency in isolating foreground elements and detecting key structural components of the galaxy with minimal pre-processing.

\begin{figure*}
    \centering
    \includegraphics[width=\textwidth]{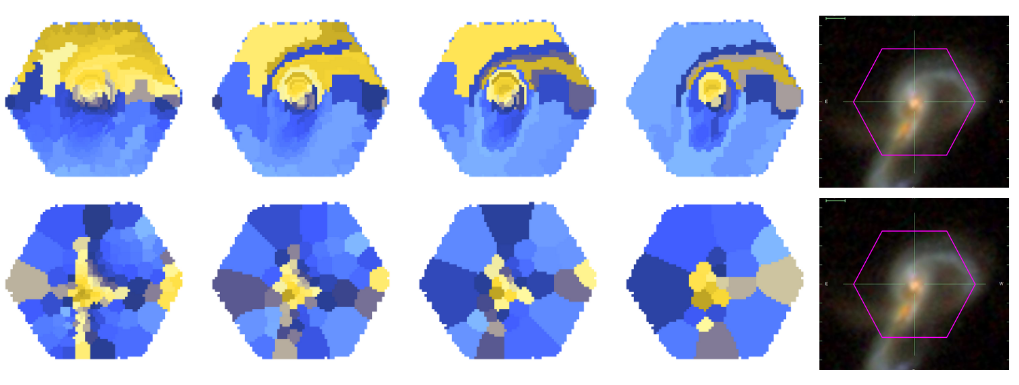}
    \caption{ \textbf{Top row}: {\sc capivara} segmentation maps for varying target SNR levels, which correspond to different numbers of components. From left to right, the typical SNR values are approximately 50, 100, 150, and 200. \textbf{Bottom row}: The Voronoi binning output for the same target SNR levels. In both rows, the rightmost panel displays the MaNGA galaxy 7443-12703 for visual comparison. 
    }
    \label{fig:cap_vor_SN}
\end{figure*}

 In Fig. \ref{fig:cap_vor_SN} we compare our results with Voronoi tessellation at various granularity levels. We run {\sc vorbin} across a range of target signal-to-noise (S/N) ratios: {50, 100, 150, 200, 250}. Each target S/N corresponds inversely to the number of resulting regions—the higher the target S/N, the fewer the regions generated. Subsequently, we run capivara with a comparable number of regions to facilitate a meaningful comparison between the two segmentation methods.  {\sc capivara} preserves an aesthetically pleasing visual resemblance to the composite SDSS image. This visual coherence is desirable, as our objective extends beyond mere S/N enhancement—we aim to group pixels into physically meaningful regions reflective of intrinsic galactic structures.

\begin{figure*}
    \centering
    \includegraphics[width=\linewidth]{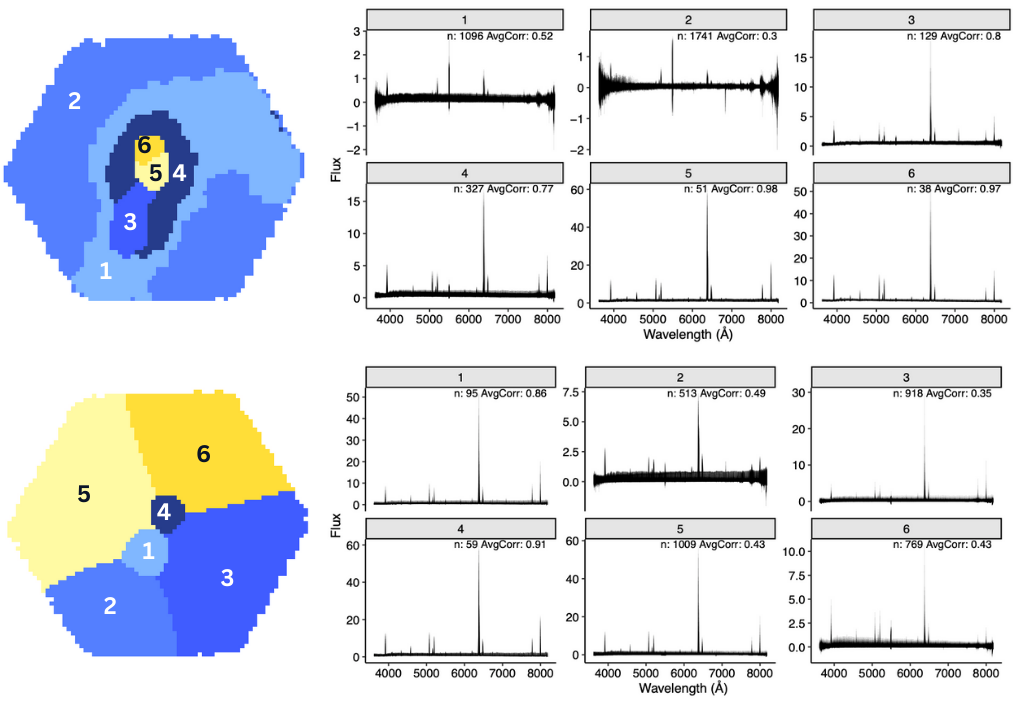}
    \caption{ Comparison of segmentation approaches: the top panel shows results from {\sc capivara}, while the bottom panel displays Voronoi binning via the \texttt{vorbin} package. In both cases, the left panels illustrate the spatial distribution of six distinct regions, and the right panels show all individual spectra within each group (rather than a single median spectrum). Each group is annotated with the number of spaxels (\texttt{n}) and the average pairwise correlation (\texttt{AvgCorr}), which quantifies the internal spectral similarity. This comparison highlights how {\sc capivara} preserves spectral-shape diversity within clusters, in contrast to Voronoi binning, which aggregates spaxels to achieve a target S/N, potentially mixing different spectral shapes.
    }
    \label{fig:capivara_reg_group}
\end{figure*}

 Finally, to provide both a qualitative and quantitative impression of our model compared to standard Voronoi binning, we conducted an illustrative experiment with six groups. This choice, made without loss of generality, serves merely to convey the intuitive qualitative differences between the two strategies. A key desideratum for any clustering approach is to enhance the SNR while preserving spectral similarity. Increasing the SNR by combining highly dissimilar spectra may indeed improve noise characteristics; however, this also dilutes the intrinsic physical properties of the region\footnote{For example, reaching the target SNR may require summing a large number of spaxels, but this predominantly accumulates noise, thereby merging regions that are not physically connected.}.
Figure~\ref{fig:capivara_reg_group} illustrates this point. In both the top and bottom maps, each color-coded region corresponds to a cluster of pixels found by {\sc capivara} and  {\sc vorbin}, respectively. On the right, we show all spectra associated with a given region, annotated with two key metrics: \texttt{n}, the number of pixels in the cluster, and the average pairwise correlation among all spectra within that cluster (\texttt{AvgCorr}). Higher values of \texttt{AvgCorr} indicate more homogeneous spectral shapes, reflecting the internal consistency of each cluster. By comparing the left and right panels, one can observe both the spatial distribution of each cluster and the degree to which those pixels share a common spectral signature—highlighting {\sc capivara}’s ability to preserve shape information throughout the data.

We note that this initial analysis should be taken as a proof of concept since we are using the entire wavelength range of the spectra for clustering. For case-specific applications, we advise selecting specific ranges of the spectra to achieve a more domain-specific segmentation. Below, we show a continuum and emission line analyses to compare the results before and after decomposition.

\subsection{Stellar continuum classification}
\label{sect:continuum}


 Examining galaxies beyond the Local Group presents the challenge of not resolving individual stars, which necessitates modeling a galaxy’s integrated light (e.g., colors or full spectrum) as a combination of simple stellar populations \citep[SSPs; see, e.g.,][]{cg11, lopez2017stellar, bittner+20, pernet2024mg}. These combinations are compared with the galaxy’s observed emission, and inversion techniques are employed to infer its physical properties. In many cases, this process begins with photometric decomposition applied to broad-band galaxy images.

 Spectral fitting, on the other hand, involves matching observed spectra with SSP models while accounting for physical effects such as dust attenuation and Doppler shifts. A common approach is the inversion technique, in which a linear combination of SSP models is fitted to the observed spectrum, incorporating processes like reddening and stellar velocity dispersion. The best fit is typically determined by minimizing the $\chi^2$ statistic \citep{Walcher2011}. Once the best-fit template is identified, the galaxy’s physical properties—such as age, metallicity, and star formation history—can be inferred.
 The star formation history can be deduced from the light contribution of stellar population models in each age bin to the best-fit template. By averaging the contributions of the different SSPs, we can also infer integrated properties such as light- and mass-weighted ages and metallicities. Prior to the fitting, spurious features and emission lines are often masked to ensure a more accurate study on the stellar continuum.
\par
We employed the {\sc starlight} code \citep{CF+04,CF+05}, which fits observed spectra using a model library that accounts for kinematics and dust reddening. We utilized the E-MILES library of SSP models \citep{Vazdekis+16}, based on a revised \citet{Kroupa01} initial mass function and BaSTI isochrones \citep{Pietrinferni+04,Pietrinferni+06}. This library spans from 1680 to 50,000~\r{A}, with an optical resolution of FWHM = 2.5~\r{A}. {\sc starlight} has been rigorously tested across multiple IFS galaxy surveys; for an application to MaNGA data, see \citet{Mallmann+18}. Since {\sc capivara} outputs are in FITS format, any spectral fitting method can easily be applied to extract stellar and ionized-gas properties from our galaxy structures.

\begin{figure}
    \centering
    \includegraphics[width=\columnwidth]{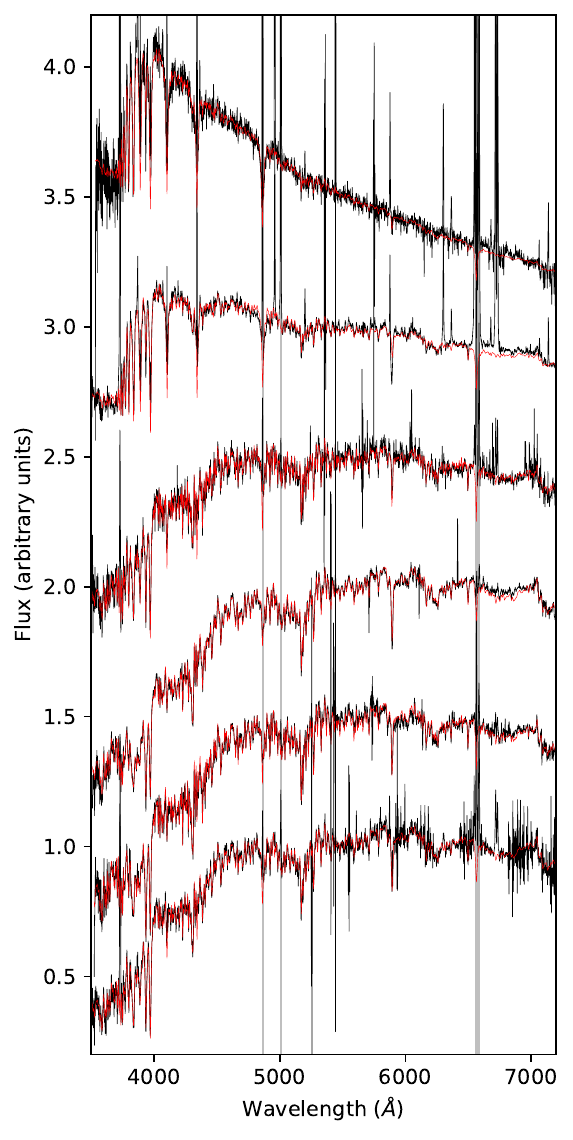}
    \caption{Observed (black) and modeled (red) integrated spectra of our sample galaxies are shown. Grey-shaded regions correspond to the areas integrated to construct the BPT diagram. For better visualization, we present these spectra in order of inclination. From top to bottom: 10224-6104, 7443-12703, 11749-12701, 9039-9102, 8443-6102, and 8135-12701.}
    \label{fig:spec}
\end{figure}

To expedite our analysis, we restricted our library to 10 representative ages\footnote{0.03, 0.06, 0.1, 0.3, 0.6, 1.0, 3.0, 6.0, 10.0, and 13.0 Gyrs} and 3 representative metallicities\footnote{Z = 0.2, 1, and 2 $Z_{\odot}$}. We also included a featureless continuum (FC) with F$_\nu \propto \nu^{-1.5}$ to account for potential AGN emission, as well as very young ($<$ 5 Myr) SSPs, which are indistinguishable from AGN-type continua \citep[][]{Koski1978,StorchiBergmann+00,Riffel+09}. Caution is therefore advised when interpreting this component. Figure~\ref{fig:spec} shows the integrated observed spectra of each galaxy in black, with the corresponding {\sc starlight} model spectra in red\footnote{For galaxy 10224-6104, the adopted SSP base may not fully capture its stellar populations. Incorporating younger populations reveals clear signs of recent star formation. See \url{https://manga.if.ufrgs.br/explorer/215} for an interactive visualization}.

Since small differences in stellar population composition are often obscured by noise in real data \citep{CF+04,CF+05,Riffel+09}, we grouped the SSP models into three age bins: young (xy: $\leq$ 100 Myr), intermediate-age (xi: 0.1–2.0 Gyr), and old (xo: 2–13 Gyr). This categorization reflects the dominant contributors to the integrated light in each bin: blue, hot stars in the young population; evolved giants, particularly TP-AGB stars, in the intermediate-age range; and cooler, red stars in the old population.
\par

\begin{figure*}
    \centering
    \begin{tabular}{c}
        \includegraphics[width=0.8\textwidth]{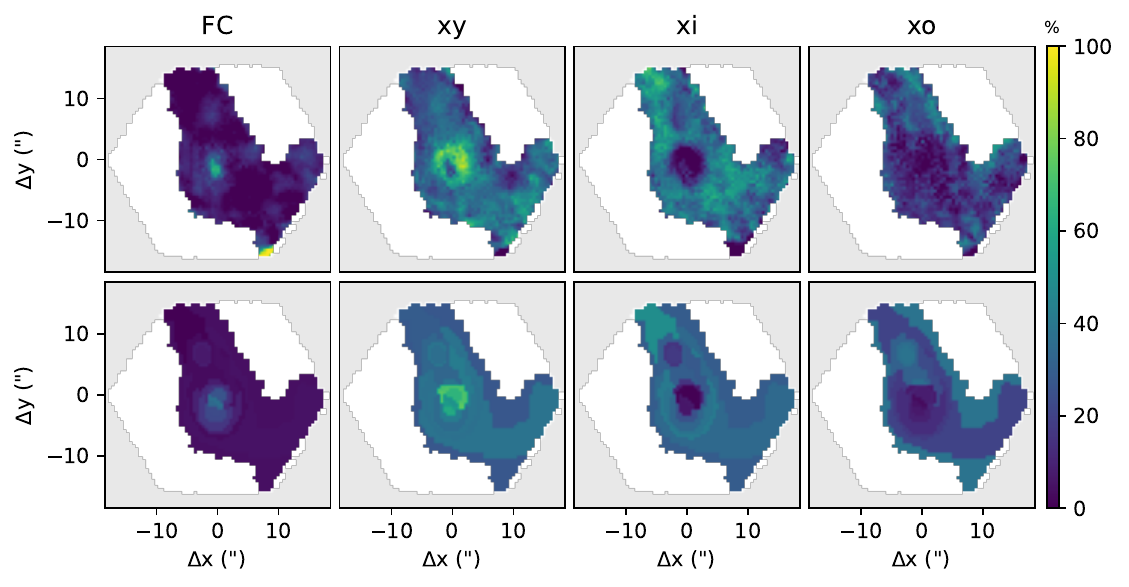} \\
        \includegraphics[width=0.8\textwidth]{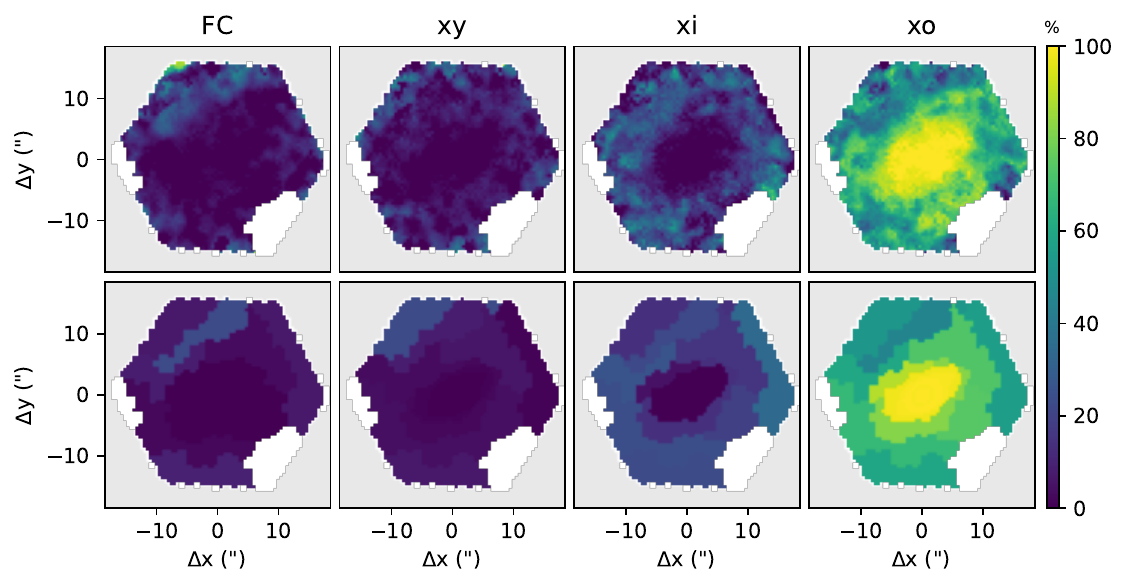} \\
    \end{tabular}
    \caption{Stellar population demographics for 7443-12703 (top) and 8135-12701 (bottom). \textbf{Top row} in each set: Regions derived from the original data cube. \textbf{Bottom row} in each set: Results from the decomposed data cube using {\sc capivara}. The categories, from left to right, are featureless continuum (FC), young (xy: $\leq$ 100 Myr),  intermediate-age (xi: 0.1-2.0~Gyr) and old (xo: 2-13 Gyr) stellar populations. The color bar indicates the fractional contribution of each age bin to the total light, with brighter colors representing higher contributions.}
    \label{fig:stpop_combined}
\end{figure*}

\begin{figure*}
    \centering
    \begin{tabular}{c}
        \includegraphics[width=0.8\textwidth]{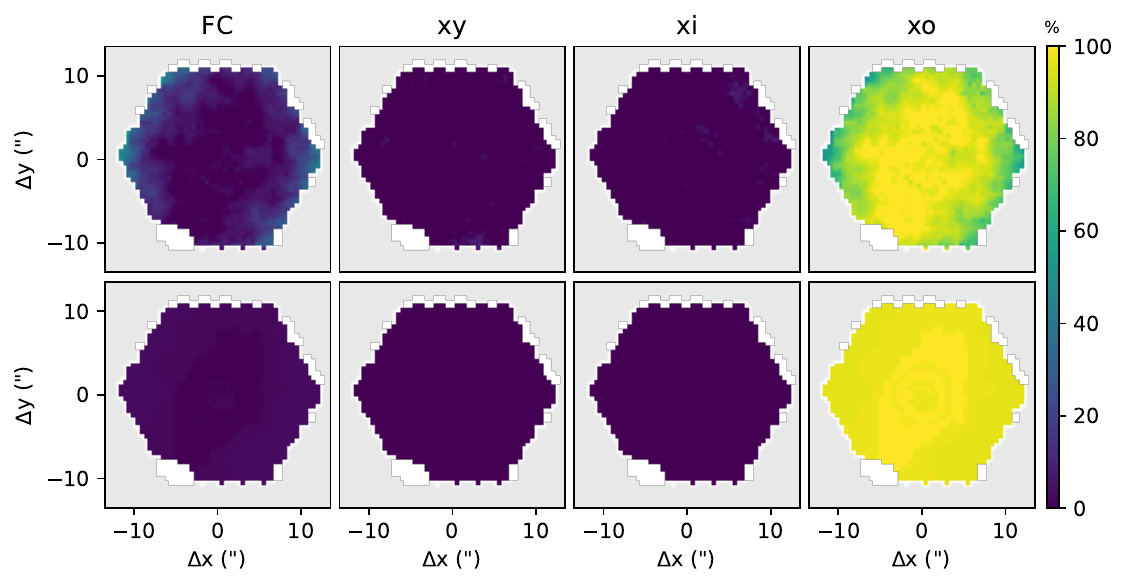} \\
        \includegraphics[width=0.8\textwidth]{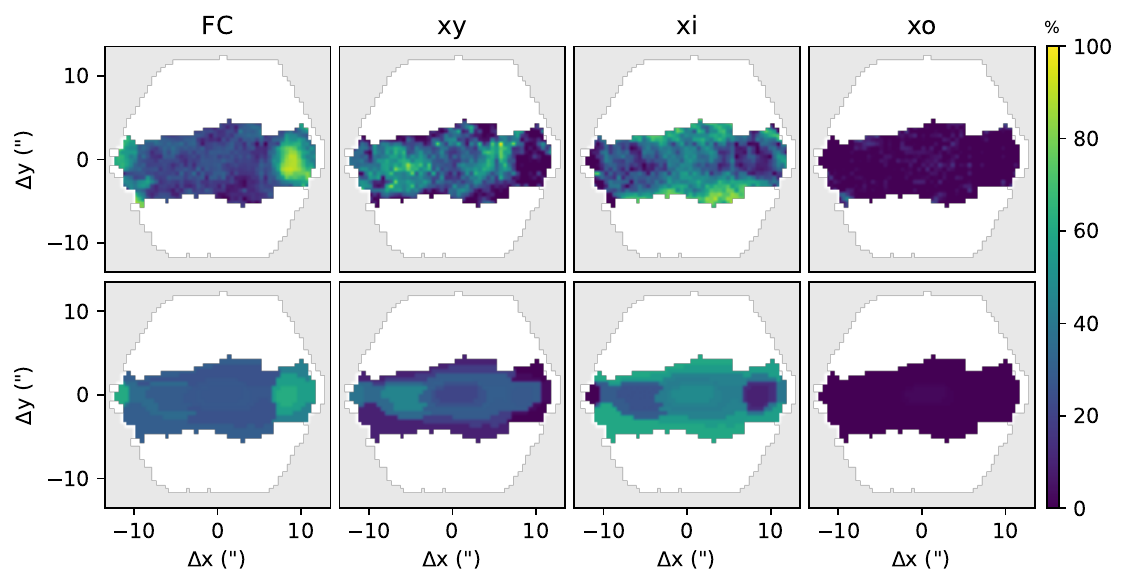} \\
        \includegraphics[width=0.8\textwidth]{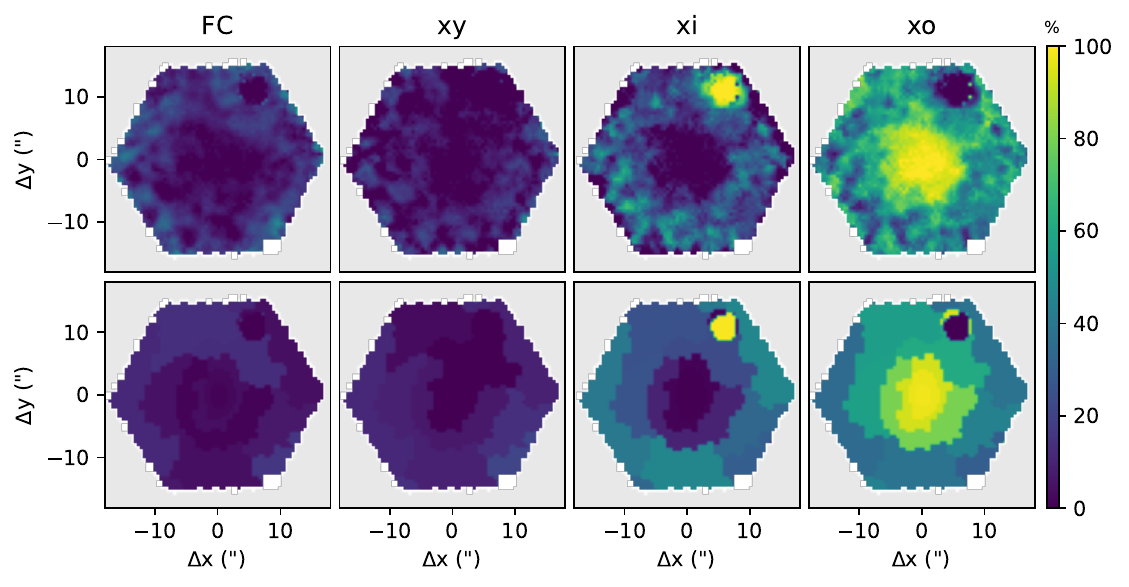} \\
    \end{tabular}
    \caption{Stellar population demographics for 8443-6102 (top), 10224-6104 (middle), and 11749-12701 (bottom). \textbf{Top row} in each set: Regions derived from the original data cube. \textbf{Bottom row} in each set: Results from the decomposed data cube using {\sc capivara}. The categories, from left to right, are featureless continuum (FC), young (xy: $\leq$ 100 Myr),  intermediate-age (xi: 0.1-2.0~Gyr) and old (xo: 2-13 Gyr) stellar populations. The color bar indicates the fractional contribution of each age bin to the total light, with brighter colors representing higher contributions.}
    \label{fig:stpop_combined2}
\end{figure*}

In Figs. \ref{fig:stpop_combined} to \ref{fig:stpop_combined2}, we present FC and condensed stellar population vectors in light fraction. The top row shows the results before decomposition, while the bottom row displays the results after decomposition. Our method effectively identifies the main regions in each galaxy, preserving their distinct features. For the galaxy 7443-12703, our code successfully identified two regions with younger stellar populations and stronger FC contributions:  one concentrated in the nucleus and another located 4\farcs0 East and 8\farcs0 North of the nucleus. Additionally, it accurately grouped the semi-ring structure observed in the $xy$ age bin panel. In the galaxy 8135-12701, our method identified the bulge and disk regions, with the bulge dominated by an old stellar population (SP) and the disk primarily by intermediate-age SPs. 
Additionally, our method identified a spiral arm with elevated FC contributions compared to the rest of the galaxy, clearly visible in the upper left region of the 8135-12701 FC panel. This apparent FC excess is likely a consequence of fitting degeneracies, where {\sc starlight} is driven to compensate for limitations in the base set’s representation of the youngest populations. The fitting process is influenced by reddening and flux calibration effects, which often force unphysical negative values that artificially shift the fits toward bluer solutions \citep[see, e.g.,][for a more detailed discussion]{Riffel2022MNRAS}.
\par
Galaxy 8443-6102, the only early-type galaxy in our sample, exhibits the simplest stellar population structure. It consists of an older central region and outer areas within our Field of View (FoV), where a small FC contribution ($\sim$10\%) was detected. Additionally, a subtle ring-like structure with a slightly younger stellar population is present. Our method accurately identified all these regions, particularly the faint ring-like structure, which is barely visible in the old age bin panel. Unlike 8443-6102, 10224-6104 displays a complex structure with three distinct light centers and an uneven stellar population. The rightmost light source is dominated by an FC-type continuum, the leftmost by a young stellar population, and the central, brightest source by an intermediate-age stellar population. {\sc capivara} was able to detect each light source and group their influence regions together, highlighting these differences. Additionally, a fourth region with a different continuum composition was detected in the left edge of our FoV, showing a slightly higher FC contribution if compared to its surroundings.

Our final object, 11749-12701, is especially notable because, despite its simple stellar population—featuring an old bulge and younger outskirts — a field star is also present. Using our approach, we automatically identified the galaxy's structures and separated the field star from the rest of the FoV.

\subsection{Emission lines classification}

\begin{figure}
    \includegraphics[width=0.95\columnwidth]{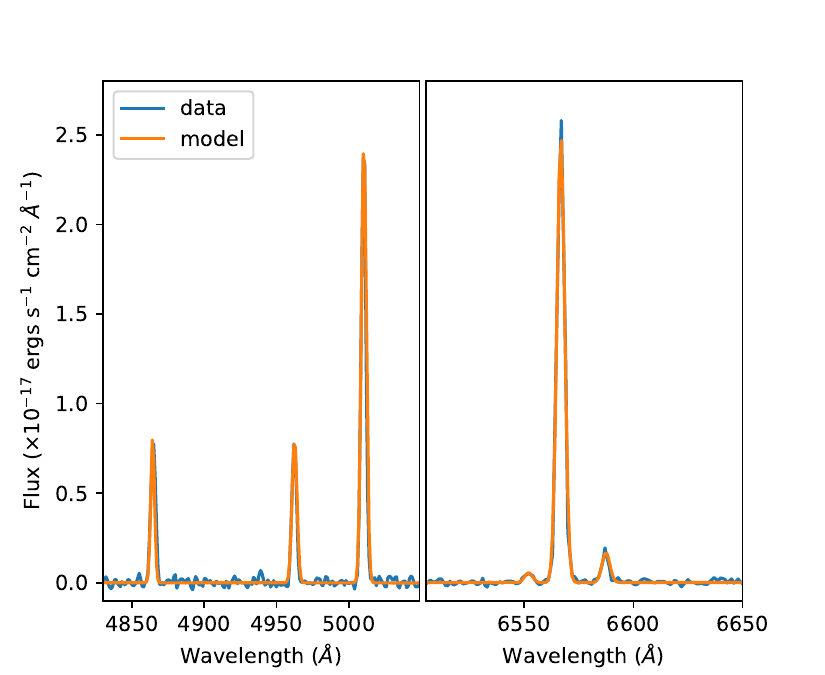}
    \caption{Fitting example for the (10\farcs0, 0\farcs0) spaxel of 10224-6104. Each emission line was fitted with a single Gaussian profile.}
    \label{fig:fit}
\end{figure}

After subtracting the stellar spectra derived in section~\ref{sect:continuum}, we measured the flux of the four main emission lines ({H$\alpha$, H$\beta$, [N{\sc ii}]$\lambda$6583\r{A} and [O{\sc iii}]$\lambda$5007\r{A}}) required to build the Baldwin-Phillips-Terlevich diagram \citep[BPT;][]{Baldwin1981,VeilleuxOsterbrock1987,rola1997,kewley+01,kauffmann+03,Stasinska2007,Schawinski2010}. This analysis was conducted by fitting a Gaussian function to each emission line using the {\sc ifscube} code \citep{Ruschel-Dutra+21}. {\sc ifscube} fits multiple Gaussian or Gauss-Hermite profiles, with or without constraints, and supports pixel-by-pixel uncertainties, weights, flags, and pseudo-continuum fitting. The best fit is determined by minimizing the $\chi^2$, the free parameters being the amplitude, radial velocity, and velocity dispersion of each component. Our sample galaxies exhibited no signs of multiple components per emission line, and a single narrow Gaussian was sufficient. Given that similar ionization potentials of H$\alpha$, H$\beta$ and [N$_{\rm II}$] have similar ionization potentials, their kinematic properties were fixed, while [O$_{\rm III}$] was allowed independent radial velocity and velocity dispersion. Fig~\ref{fig:fit}, provides an example fit for galaxy 10224-6104 at (10\farcs0, 0\farcs0).

\begin{figure*}
    \centering
    \begin{tabular}{c}
        \includegraphics[width=0.85\textwidth]{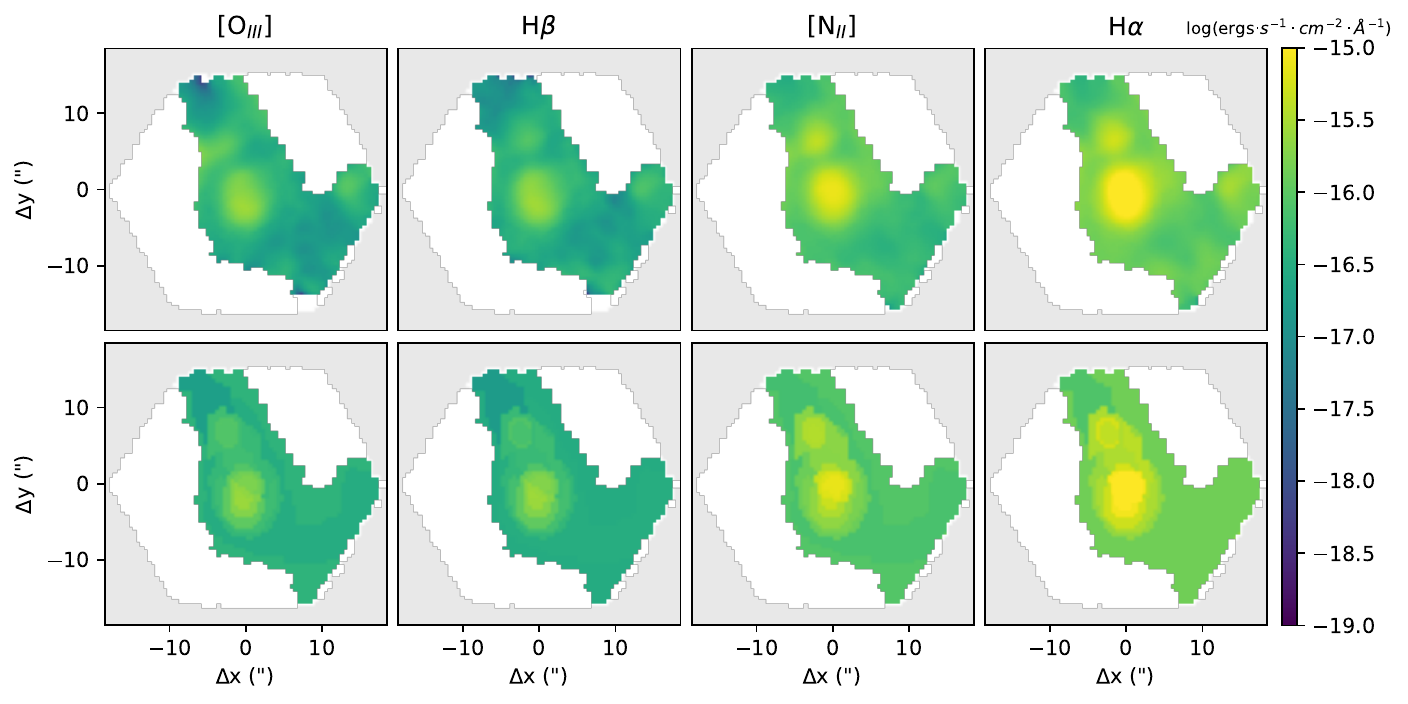} \\
        \includegraphics[width=0.85\textwidth]{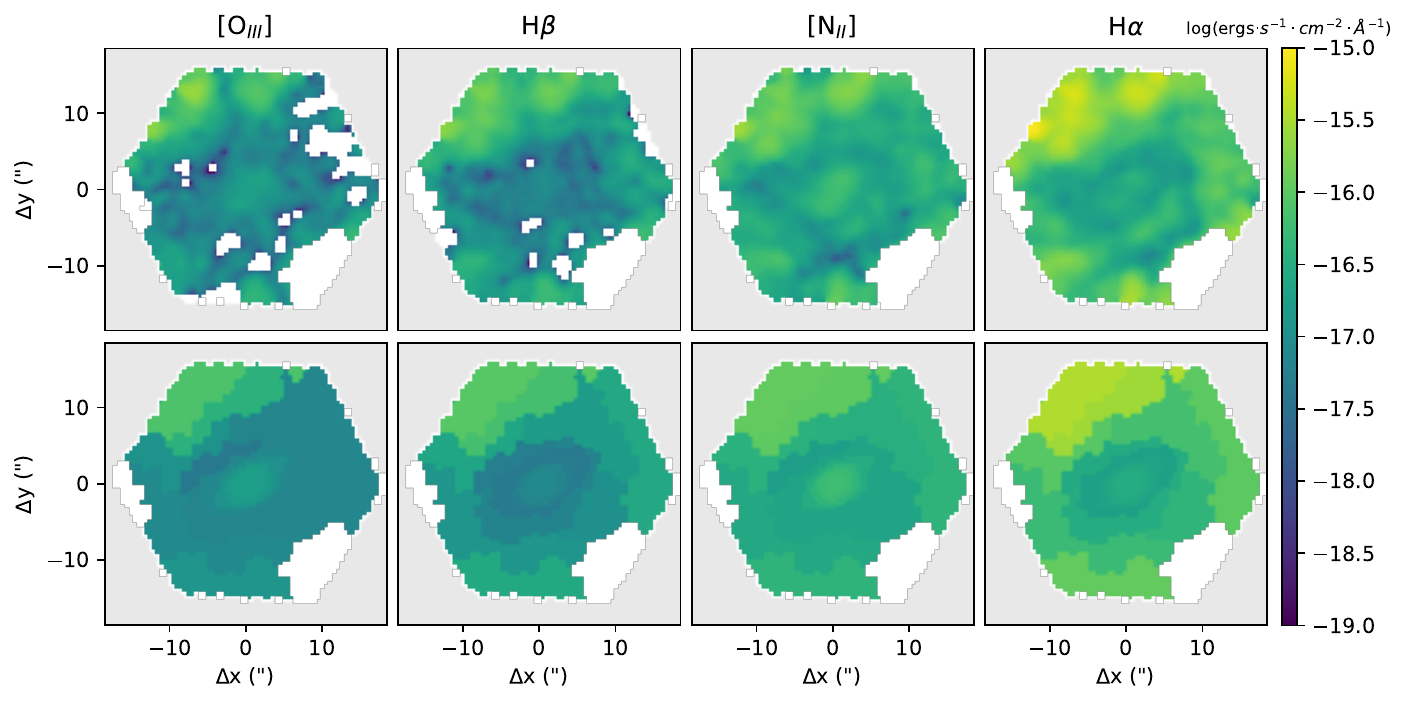} \\
    \end{tabular}
    \caption{Logarithmic fluxes of the four main emission lines for 7443-12703 (top) and 8135-12701 (bottom), in ergs$\cdot$s$^{-1} \cdot$cm$^{-2} \cdot$\r{A}$^{-1}$. \textbf{Top row} in each set: Fluxes measured in the original datacube. \textbf{Bottom row} in each set: Fluxes measured in the decomposed datacube using {\sc capivara}.}
    \label{fig:emlin_combined}
\end{figure*}

\begin{figure*}
    \centering
    \begin{tabular}{c}
        \includegraphics[width=0.8\textwidth]{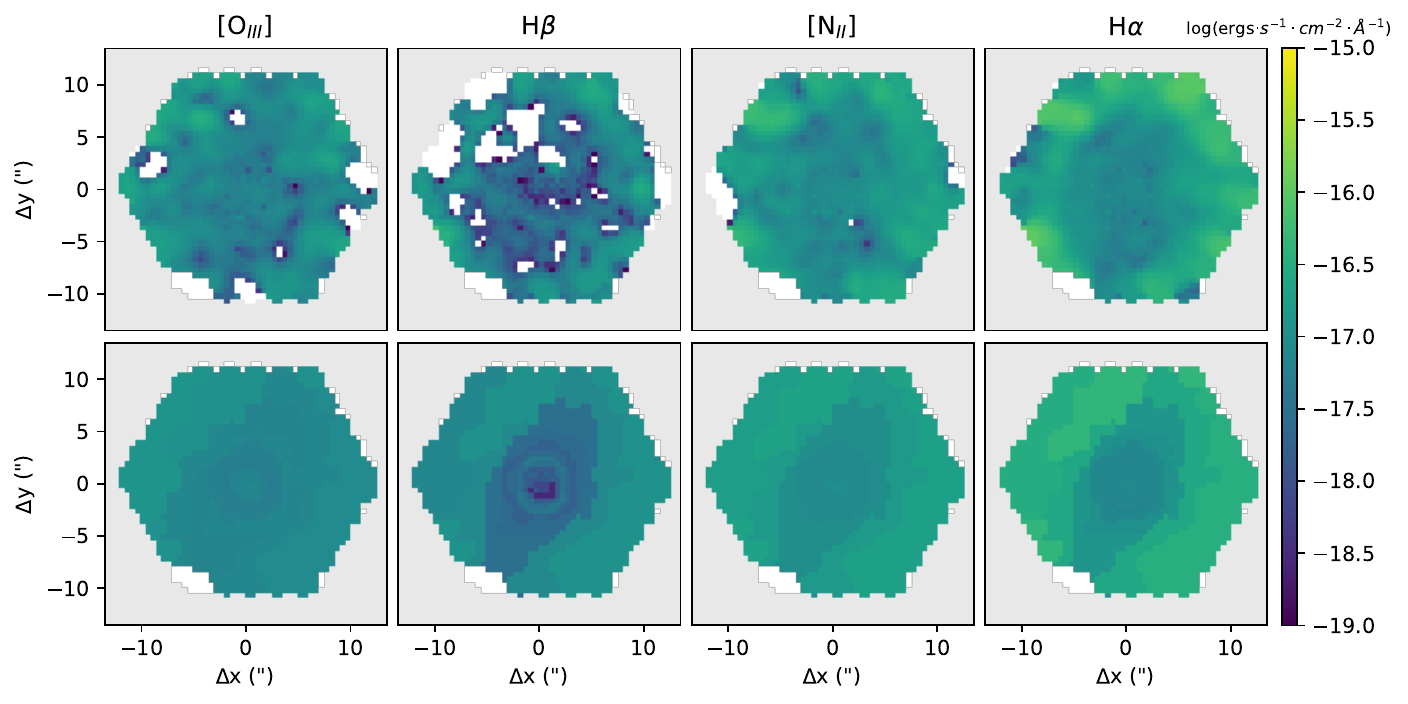} \\
        \includegraphics[width=0.8\textwidth]{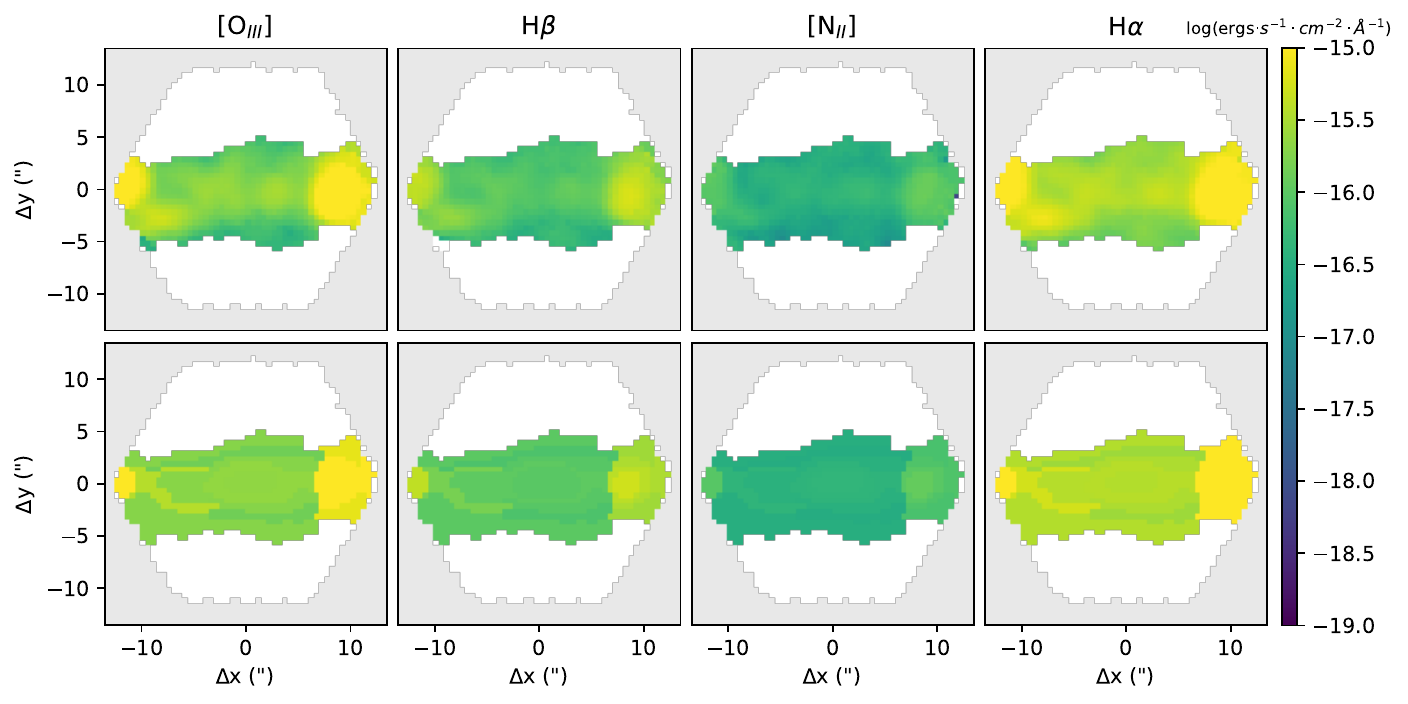} \\
        \includegraphics[width=0.8\textwidth]{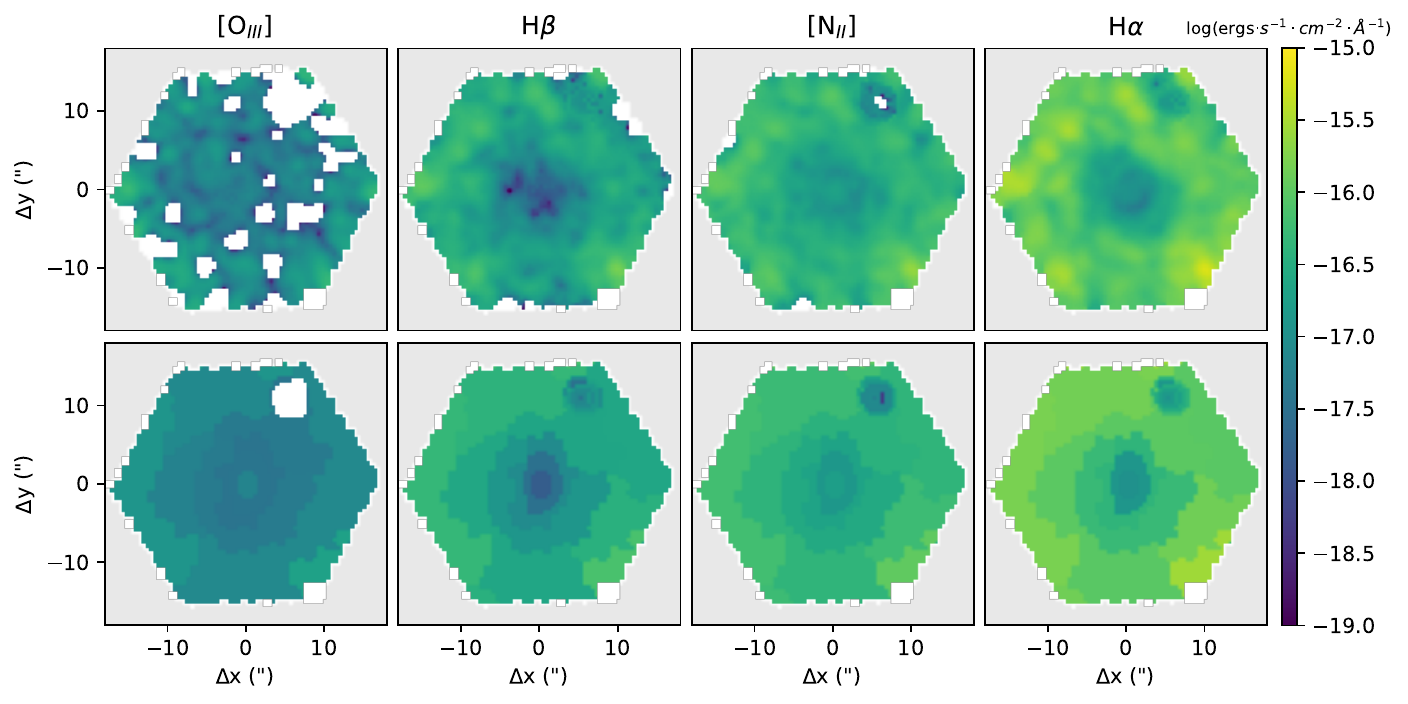} \\
    \end{tabular}
    \caption{Logarithmic fluxes of the four main emission lines for 8443-6102 (top), 10224-6104 (middle), and 11749-12701 (bottom), in ergs$\cdot$s$^{-1} \cdot$cm$^{-2} \cdot$\r{A}$^{-1}$. \textbf{Top row} in each set: Fluxes measured in the original datacube. \textbf{Bottom row} in each set: Fluxes measured in the decomposed datacube using {\sc capivara}.}
    \label{fig:emlin_combined3}
\end{figure*}

The H$\alpha$, H$\beta$, [N{\sc ii}], and [O{\sc iii}] maps are displayed in Figures~\ref{fig:emlin_combined} and \ref{fig:emlin_combined3}. From left to right, the panels depict the fluxes for [\ion{O}{iii}], H$\beta$, [\ion{N}{ii}], and H$\alpha$. The top row presents the fluxes measured from the original data cube, while the bottom row shows the measurements after running {\sc capivara}.
Similar to the stellar continuum, our method effectively identifies the main emission-line regions in each object. However, the improvement in S/N is much more pronounced in the emission lines. In many regions of the original datacube, these features are not properly measured due to not meeting the S/N threshold. This issue is resolved in the decomposed datacube, where our model not only enhances the S/N and correctly groups the regions but also ensures more consistent and accurate measurements of these features, without merging regions that are not physically meaningful from a spectral perspective.

In our sample, regions identified by their distinct continuum properties also exhibit differences in emission line properties, so the figures typically highlight the same regions discussed in Section~\ref{sect:continuum}. However, even when two regions share similar continuum properties but differ in their emission lines, or vice versa, our approach should be able separate them effectively. Moreover, depending on the spectral features being studied, different techniques can be applied. For instance, when focusing on the stellar continuum, our method can be performed on the original datacube since there are significantly more continuum points than emission line points. Alternatively, if the study targets emission lines, one can analyze the derivative of the original datacube. In regions where the continuum dominates, the derivative will be close to zero, whereas in areas dominated by complex emission lines, it will deviate significantly from zero.

In this way, {\sc capivara} can be tailored to focus on specific features of interest, classifying different regions of the galaxy based on those chosen features. Since this paper focuses on the methodology rather than specific astrophysical cases, which will be addressed in follow-up papers, we performed both continuum and emission line analyses based on the initial decomposition into 20 distinct regions. Even with this straightforward approach, the characteristics of the original object were successfully recovered, showcasing the simplicity and efficacy of our method.

\subsection{Continuum and emission line spectral classification}

Although the same galaxy can be classified into different categories based on ad hoc criteria or specific predetermined wavelength windows, a general sanity check for {\sc capivara} is to ensure that its spectral segmentation is qualitatively consistent with traditional continuum and emission line-based classifications.
To achieve this, we classified each representative spectrum based on both continuum features and emission line characteristics.
\par

We divided our sample into five classes according to the dominant stellar populations in each object, presented here in the order of appearance: AGN, star-forming (SF), post-starburst (PS), mixed (MIX), and quenched (QUENCH). The AGN class includes objects dominated by a FC component. The SF class consists of objects where simple stellar populations (SSPs) younger than 100 Myr dominate the emission. The PS class is defined by objects in which SSPs between 100 and 2000 Myr dominate. In the MIX class, SSPs older than 2 Gyr contribute less than 40\% of the luminosity, while in the QUENCH class, SSPs older than 2 Gyr contribute more than 40\% of the luminosity.
\par
The BPT diagram classifies galaxies into five distinct categories based on their emission-line fluxes: inactive (INACT), star-forming (SF), composite or transition (TRANSIT), low-ionization nuclear emission-line region (LINER), and Seyfert galaxies. Here, we utilize the BPT diagram based on the flux ratios of H$\beta$, [OIII] $\lambda$5007, H$\alpha$, and [NII] $\lambda$6583, plotting galaxies in a parameter space defined by $\log (\mathrm{[NII]}/\mathrm{H}\alpha)$ on the x-axis and $\log (\mathrm{[OIII]}/\mathrm{H}\beta)$ on the y-axis. The thresholds between these line ratios, as defined by \citet{kewley+01} and \citet{kauffmann+03}, are used to differentiate between these classes.
Inactive galaxies are identified by [O{\sc iii}] emission below 3-$\sigma$ of the continuum signal-to-noise ratio (S/N). Star-forming galaxies, primarily influenced by photoionization from massive young hot stars, exhibit low values of both [N{\sc ii}]/H$\alpha$ and [O{\sc iii}]/H$\beta$, occupying the left wing locus of the BPT diagram. In contrast, AGN-dominated galaxies, including LINERs and Seyferts, lie on the opposite side. LINERs display high [N{\sc ii}]/H$\alpha$ but low [O{\sc iii}]/H$\beta$ ratios, with their ionization source being uncertain, possibly due to nuclear activity or evolved stellar populations. Seyferts show high values in both [N{\sc ii}]/H$\alpha$ and [O{\sc iii}]/H$\beta$ ratios. Finally, the transition class falls in the zone between star-forming and AGN objects, demarcated by the theoretical extreme starburst line proposed by \citet{kewley+01} and the empirical starburst line by \citet{kauffmann+03}.
\par
\begin{figure*}
    \centering
    \begin{tabular}{c}
        \includegraphics[width=0.75\textwidth]{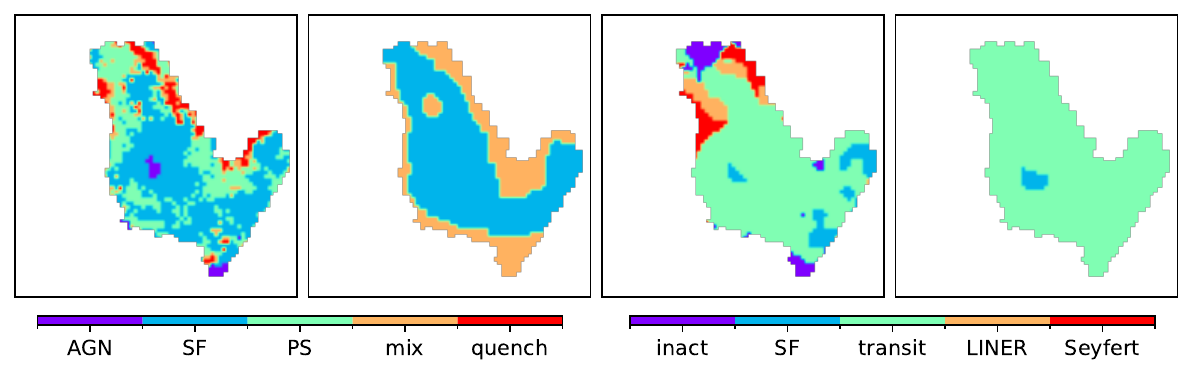} \\
        \includegraphics[width=0.75\textwidth]{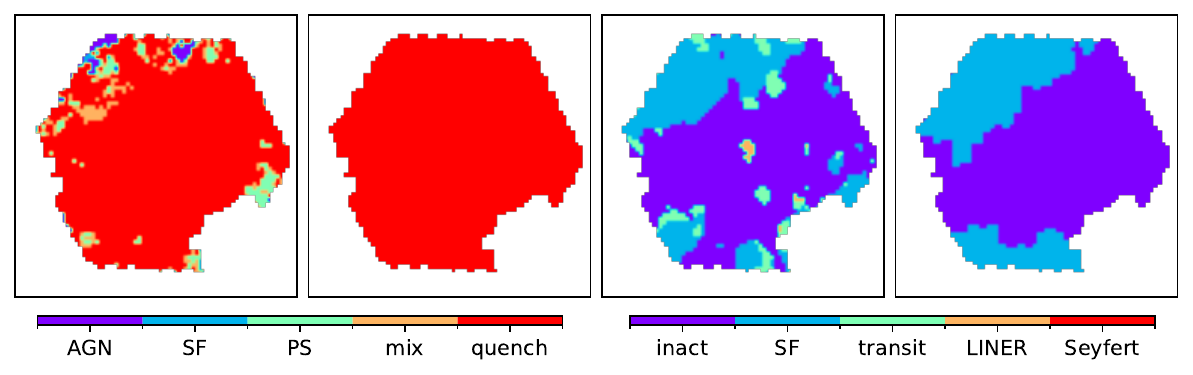} \\
        \includegraphics[width=0.75\textwidth]{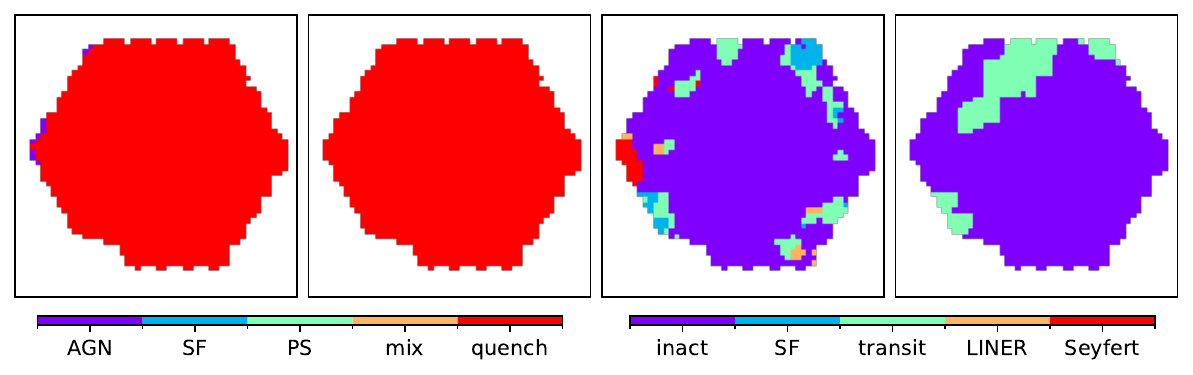} \\
        \includegraphics[width=0.75\textwidth]{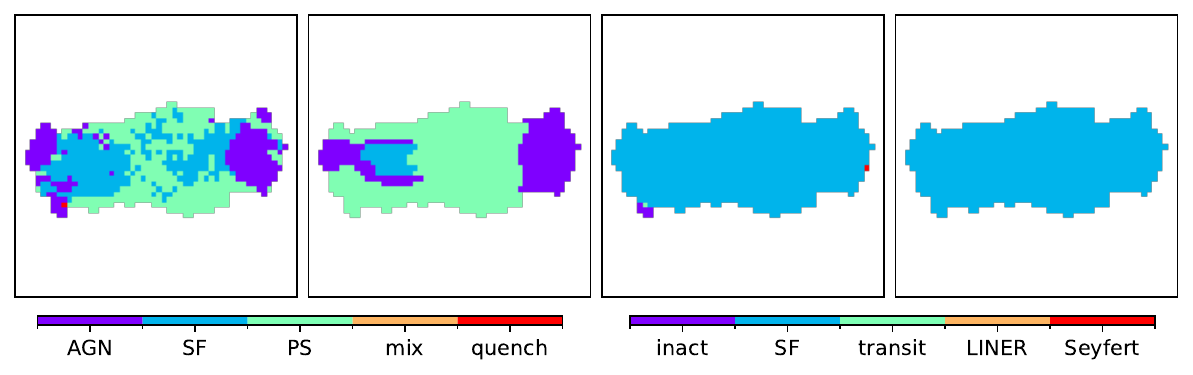} \\
        \includegraphics[width=0.75\textwidth]{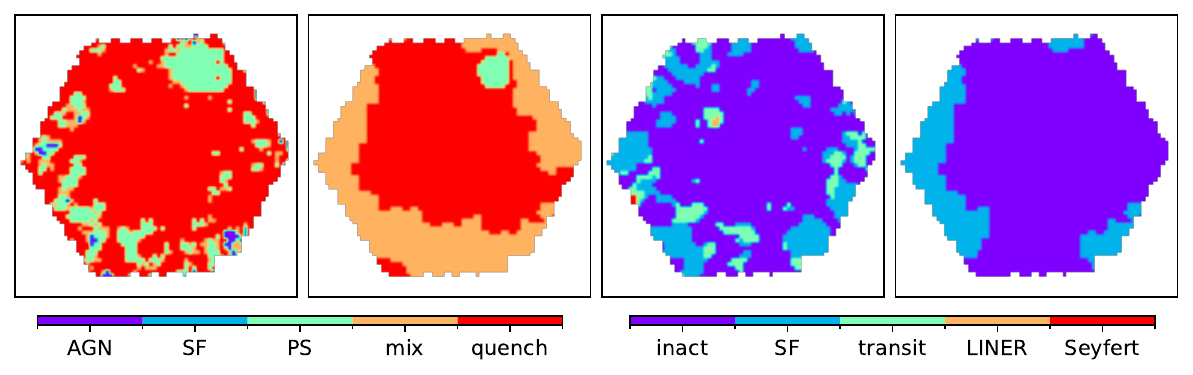} \\
    \end{tabular}
    \caption{Spectral classification maps for five regions: 7443-12703 (top), 8135-12701 (second from top), 8443-6102 (middle), 10224-6104 (second from bottom), and 11749-12701 (bottom). The first two panels on the left display the maps for the continuum classes, while the two panels on the right present the classification based on emission lines. In each case, the left panels show the analysis on the original datacube, and the right panels show the analysis in the decomposed datacube using {\sc capivara}.}
    \label{fig:combined_classes5}
\end{figure*}

 Figure~\ref{fig:combined_classes5} shows the classification maps for our five galaxies. The first two panels on the left in each row display the maps for the continuum classes, while the two on the right present the classifications based on emission lines. In both cases, the original datacube is shown first, followed by the segmented version.  Before the {\sc capivara} segmentation, many galaxies exhibit low S/N regions where classifications fluctuate across adjacent spaxels, often within areas smaller than the PSF. These fluctuations are not physically meaningful, as coherent structures should span spatial scales larger than the PSF. While our method does not explicitly enforce spatial coherence, it preserves spectral similarity across regions, which naturally leads to more spatially coherent segmentations. This, in turn, mitigates the likelihood of merging physically distinct regions. Nevertheless, in some cases, adjacent regions with subtle spectral differences may still be grouped together, leading to the dilution or suppression of less dominant classes. This limitation can be alleviated by increasing the number of classes, though the optimal choice is likely galaxy-dependent and best addressed in dedicated applications. Even so, the segmented regions often align with physically distinct components—whether driven by gas excitation, stellar populations, or a combination of both.

\section{Conclusions}
\label{sec:voronoi}

Traditional methods for analyzing astronomical data cubes often struggle to extract full signals from noisy backgrounds, particularly when astrophysical or instrumental correlations between nearby measurements are underestimated. To address these challenges, efforts have been made to incorporate spatial information, such as global parametric models for physical distributions in kinematic studies \citep[e.g.,][]{Krajnovic06,watkins+13} and automated spatial segmentation \citep{batman}. One common approach to improve signal-to-noise ratio (S/N) is through adaptive binning techniques like Voronoi tessellation \citep{Cappellari&Copin03}, which groups spaxels into contiguous regions for averaged measurements \citep{Sanders01,Diehl06,Sanchez16}. While these techniques enhance statistical confidence, they often do so at the expense of spatial resolution.

Our method, however, takes a different approach. Primarily designed to segment data into regions with distinct physical properties, it naturally enhances the S/N as a secondary benefit. In contrast to Voronoi binning—which focuses on data quality by merging regions that may have different physical characteristics—our approach ensures that only regions with similar properties are grouped. In our current analysis, segmenting the data cube into 20 distinct regions results in a five-fold improvement in S/N compared to individual spectra.\footnote{This estimate was derived from the continuum between 5500 and 5600 \r{A}, chosen for its minimal stellar absorption and emission lines.}

Building on this, we introduce \textsc{capivara}, a novel spectral-based segmentation method for IFU data cubes. Designed specifically to enhance the study of structural properties in galaxies, \textsc{capivara} employs hierarchical clustering in the spectral domain, grouping similar spectra to improve S/N without compromising astrophysical similarity among regions.

 As a caveat, the quality of the segmentation produced by \textsc{capivara} ultimately depends on the spectral features present in the data. This is not a limitation of the method itself, but rather a reflection of the information content available to the algorithm. Users should consider pre-processing the spectra for specific needs, emphasizing or isolating regions of interest—such as prominent emission or absorption features—or adopting more physically motivated distance metrics, depending on the scientific goal. Another current limitation of our package is the lack of correction for internal kinematics, such as velocity shifts across the galaxy. While this does not appear to significantly impact the results for the systems analyzed—since spectral variations from ionization or stellar populations tend to dominate—it may become more relevant in galaxies with strong velocity gradients. Future incarnations of the code may address this effect through case-by-case implementations that incorporate kinematic effects into the clustering process. 

In a demonstration using five selected MaNGA galaxies, \textsc{capivara} identified and grouped regions with similar physical properties. These groupings were validated against standard methods for stellar continuum and gas property analysis. By providing both segmented data cubes and integrated one-dimensional spectra for each region, \textsc{capivara} facilitates a more detailed investigation of stellar and ionized gas properties.

\section*{Acknowledgements}

R. S. de Souza acknowledges support from the São Paulo Research Foundation (FAPESP), project 2024/05315-4. L.G.D.H. acknowledges support from the National Key R\&D Program of China (2022YFF0503402) and the National Natural Science Foundation of China (NSFC, E345251001). A.C.S. acknowledges funding from the Brazilian National Council for Scientific and Technological Development (CNPq, grants 314301/2021-6 and 445231/2024-6) and the Rio Grande do Sul State Research Foundation (FAPERGS, 24/2551-0001548-5). P.C. acknowledges support from CNPq (310555/2021-3) and FAPESP (2024/05315-4). A.C.S. and R.R. also acknowledge support from the Coordination for the Improvement of Higher Education Personnel (CAPES, project 0001). R.R. further acknowledges support from CNPq (404238/2021-1, 445231/2024-6) and FAPERGS (24/2551-0001282-6), as well as from CAPES (88881.109987/2025-01). 

\section*{Data Availability}

This work uses data from the MaNGA DR17 data release, available at \url{https://www.sdss4.org/dr17/manga/}.


\bibliographystyle{mnras}
\bibliography{ref} 

\bsp	
\label{lastpage}
\end{document}